\documentclass{article}
\usepackage{graphicx}
\title{Anomalies in Quantum Mechanics: the $1/r^2$ Potential}
\author{Sidney A. Coon\\
Physics Division\\
National Science Foundation\\
4201 Wilson Blvd.\\
Arlington, VA 22230\\
and\\
Physics Department\\
New Mexico State University\\
Las Cruces, NM 88003\\
and\\
Barry R. Holstein\\
Institut f\"{u}r Kernphysik\\
Forschungszentrum J\"{u}lich\\
D-52425 J\"{u}lich, Germany\\
and\\
Department of Physics\\
University of Massachusetts\\
Amherst, MA 01003}
\begin{document}\begin{titlepage}
\maketitle
\begin{abstract}
An anomaly is said to occur when a symmetry that is valid
classically becomes broken as a result of quantization.
Although most manifestations of this phenomenon are in the context
of quantum field theory, there are at least two cases in quantum
mechanics --- the two-dimensional delta function interaction and
the $1/r^2$ potential. The former has been treated in this
journal; in this article we discuss the physics of the latter
together with experimental consequences.
\end{abstract}
\end{titlepage}
\section{Introduction}
The use of symmetry to enhance our understanding of
physical systems is well-known. An important manifestation of
this simplification is N\"{o}ther's theorem, which guarantees the
correspondence between symmetries and conservation
laws.$^{\ref{bib:noe}}$ Examples of the consequences of this
theorem include 
\begin{itemize}
\item [(a)] translation invariance $\leftrightarrow$ momentum
conservation;
\item [(b)] time translation invariance $\leftrightarrow$ energy
conservation;
\item [(c)] rotational invariance $\leftrightarrow$ angular 
momentum conservation.

\end{itemize}

These particular symmetries and conservation laws are exact. Far
more common are cases for which the invariance is only
approximate and is broken in some fashion. Nevertheless, the
symmetry still represents a useful way of analyzing the
system, and it is important to understand the mechanisms by which
this symmetry breaking can take place. Despite the many physical
situations involving symmetry violation, there exist just three
mechanisms by which this violation can take place:

\begin{itemize}

\item [(i)] Explicit symmetry breaking, wherein the breaking occurs
explicitly in the Lagrangian. Familiar examples include
particle physics, where the heavier mass of the strange quark
compared to its up, down counterparts violates the underlying SU(3)
invariance;$^{\ref{bib:eig}}$ nuclear physics, where the up-down
quark mass difference together with electromagnetic effects are
responsible for small deviations from isotopic spin
invariance;$^{\ref{bib:iso}}$ gravitational physics, where
general relativity together with small perturbations from the outer
planets lead to deviations from the underlying O(4) invariance
associated with a pure $1/r$ interaction and hence to the 
precession of Mercury's perihelion.$^{\ref{bib:grav}}$

\item [(ii)] Spontaneous (or hidden) symmetry breaking, wherein the
Lagrangian remains invariant, but the symmetry is not present in
the ground state. Familiar examples include condensed matter
physics, where the spontaneous violation of rotational invariance
by the creation of spin-correlated domains in materials such as
iron leads to the phenomenon of ferromagnetism,$^{\ref{bib:fer}}$
and the spontaneous violation of local gauge invariance by the
condensation of spin- and momentum-correlated electron pairs in
low temperature systems leads to
superconductivity;$^{\ref{bib:sup}}$ classical physics, where (as
first studied by Jacobi) the rapid rotation of a gravitationally
bound sphere leads to a lowest energy state not possessing the
expected axial invariance.$^{\ref{bib:axi}}$

\item [(iii)] Anomalous (or quantum mechanical) symmetry breaking,
wherein the symmetry is present at the classical level, but is
broken by quantization.$^{\ref{bib:jack}}$
Examples include elementary particle physics, where the two photon
decay of the neutral pion verifies the anomalous breaking of axial
SU(2) invariance,$^{\ref{bib:pi}}$ and in elementary particle
physics, where the so-called trace anomaly leads to a substantial
component of the nucleon mass being due to its gluon
substructure.$^{\ref{bib:pi}}$

\end{itemize}

Although these manifestations of explicit and spontaneous
symmetry breaking are textbook examples and are well
known and available to most physicists, the realization of
anomalous symmetry breaking is generally presented only within the
context of quantum field theory and is subsequently somewhat
inaccessible to all but the experts. Yet this inaccessibility
need not be the case. Indeed in previous contributions to this
journal, it has been shown that the anomaly is manifested in
ordinary quantum mechanics in two spatial dimensions with a 
delta function interaction.$^{\ref{bib:del}}$ However, this
example is not realized in nature. In this note we point
out that an additional example of anomalous symmetry
breaking occurs in the real world of three spatial dimensions in
the presence of a
$1/r^2$ potential and that the resultant predictions have been
experimentally verified in atomic physics. The discussion is at
the level appropriate for a graduate quantum mechanics course.

After a brief review of the 
previously mentioned two-dimensional
delta function potential, we show in Sec.~\ref{sec:anomalies} how
the $1/r^2$ interaction can be analyzed using either cutoff
regularization in both the bound state and scattering regimes. In
Sec.~\ref{sec:apply} we demonstrate how this situation can be
realized experimentally and discuss the confrontation of theory
with recent experiments. Our results are summarized in
Sec.~\ref{sec:concl}.

\section{Anomalies in Quantum Mechanics}\label{sec:anomalies}
To understand how an anomaly is realized in quantum
mechanics, we first review the partial wave formalism, in which 
the solution to the time-independent Schr\"{o}dinger equation
is expanded in Legendre polynomials
\begin{equation}
\psi(\vec{r})=\sum_\ell a_\ell P_\ell(\cos\theta){1\over r}
R_\ell(r)\, . \label{eq:jl}
\end{equation}
The radial functions $R_\ell(r)$ obey the differential
equation (henceforth we employ $\hbar=c=1$)
\begin{equation}
\left[-{1\over 2m}{d^2\over dr^2}+{\ell(\ell+1)\over
2mr^2}+V(r)\right]R_\ell(r)=ER_\ell(r)\,\,.\label{eq:sch}
\end{equation}

\subsection{Free Particle}
For a free particle, $V(r)=0$ and $E\equiv \vec{k}^2/2m$, and we
have the plane wave solution
\begin{equation}
\psi(\vec{r})=\exp(ikz)=\sum_\ell (2\ell+1)i^\ell 
j_\ell(kr)P_\ell(\cos\theta)\,\,,\label{eq:pla}
\end{equation}
that is, $a_\ell=i^\ell(2\ell+1)/k,\,R_\ell(r)=krj_\ell(kr)$
in the notation of Eq.~(\ref{eq:jl}). By using the asymptotic
behavior
\begin{equation}
j_\ell(kr)\stackrel{r\rightarrow\infty}{\longrightarrow}{1\over
kr}\sin(kr-\ell{\pi\over 2})\,\,,
\end{equation}
we can write Eq.~(\ref{eq:pla}) in the form
\begin{equation}
\label{eq:plane}
\exp(ikz)\stackrel{r\rightarrow\infty}{\longrightarrow}{1\over
2ikr}
\sum_\ell(2\ell+1)P_\ell(\cos\theta)(e^{ikr}-e^{-i(kr-\ell\pi)})\, .
\end{equation}
Equation~(\ref{eq:plane}) is a linear combination of
incoming, $e^{-ikr}$, and outgoing, $e^{ikr}$, spherical waves
with a phase shift $\ell\pi$ between them in the channel having
angular momentum $\ell$, due to the centrifugal potential
term, $\ell(\ell+1)/2mr^2$ in Eq.~(\ref{eq:sch}). The
existence of this {\it energy-independent} phase between incoming
and outgoing spherical wave components can be understood from the
invariance of Eq.~(\ref{eq:sch}) under the scale transformation
$r\rightarrow\mu r,\,k\rightarrow k/\mu$, which requires that the
solution be a function of the product $kr$, which is scale
invariant.$^{\ref{bib:rom}}$ This condition is obviously
satisfied by the spherical Bessel functions $j_\ell(kr)$, and
would be violated by the existence of an energy-dependent phase.

In the presence of a potential, the asymptotic form of the
scattering solutions to the Schr\"{o}dinger equation becomes
\begin{eqnarray}
\psi^{(+)}(\vec{r})
&&\stackrel{r\rightarrow\infty}{\longrightarrow}
 {1\over 2ikr}\sum_\ell(2\ell+1)P_\ell(\cos\theta)
[e^{i(kr+2\delta_\ell(k))}-e^{-i(kr-\ell\pi)}]\nonumber\\
{}&&=e^{ikz}+{e^{ikr}\over r}f(\theta)\,\,,
 \end{eqnarray}
where
\begin{equation}
f(\theta)=\sum_\ell(2\ell+1){e^{i2\delta_\ell(k)}-1\over
2ik}P_\ell(\cos\theta)
\end{equation}
is the scattering amplitude, and $\delta_\ell(k)$ is the scattering
phase shift introduced by the potential. Because the presence of
the potential breaks the scale invariance, the appearance of an
energy-dependent phase is permitted.

\subsection{$\delta^2(\vec{r})$ Potential}
Now consider what happens in two spatial
dimensions, for which the asymptotic form of the scattering
solution is$^{\ref{bib:rjk}}$
\begin{equation}
\psi^{(+)}(\vec{r})\stackrel{r\rightarrow\infty}{\longrightarrow}
e^{ikx}+{e^{i(kr+{\pi\over 4})}\over \sqrt{r}}f(\theta)\,\,,\label{eq:asy}
\end{equation}
with
\begin{equation}
f(\theta)=-i\sum_{n=-\infty}^\infty{e^{i2\delta_n(k)}-1\over
\sqrt{2\pi k}}e^{in\theta}\,\,.
\end{equation}
In this case if we include the potential energy
\begin{equation}
V(\vec{r})=\lambda\delta^2(\vec{r}) ,
\end{equation}
the scale invariance is maintained --- indeed, at the classical
level, there is no scattering because no particles are
deflected. However, the absence of scattering no longer holds
once the theory is quantized. Actually this anomalous
behavior should be expected because of the wave-like
nature of the particles --- an incident beam even with zero impact
parameter can sense the presence of the potential spike at
$\vec{r}=0$. Of course, this scattering will
occur only in the
$n=0$ channel, because wavefunctions with $n\neq 0$ must vanish at
the origin.

To see how the scattering arises, we examine a scattering solution.
In momentum space the Schr\"{o}dinger equation becomes
\begin{equation}
\label{eq:sc}
{1\over 2m}(\vec{p}\,^2-\vec{k}^2)\phi^{(+)}(\vec{p})
=-\lambda\psi^{(+)}(\vec{r}=0) ,
\end{equation}
where
\begin{equation}
\phi^{(+)}(\vec{p})=\! \int \!
d^2re^{-i\vec{r}\cdot\vec{p}}\psi^{(+)}(\vec{r})
\end{equation}
is the Fourier transform of the scattering wavefunction. 
 The solution to Eq.~(\ref{eq:sc}) is 
\begin{equation}
\phi^{(+)}(\vec{p})=(2\pi)^2\delta^2(\vec{p}-\vec{k})-
{2m\lambda\psi^{(+)}(\vec{r}=0)\over
\vec{p}^2-\vec{k}^2-i\epsilon}\,\,.
\end{equation}
For consistency, we require,
\begin{equation}
\psi^{(+)}(\vec{r}=0)= \! \int \! {d^2p\over
(2\pi)^2}\phi^{(+)}(\vec{p})
=1-2m\lambda\psi^{(+)}(\vec{r}=0){1\over 4\pi}\log
-{\Lambda^2\over k^2},
\end{equation}
where we have regulated the otherwise divergent momentum space
integration by introducing a cutoff parameter $\Lambda$. Then
\begin{equation}
\psi^{(+)}(\vec{r}=0)={1\over 1+(\lambda m/
2\pi)\bigl(\log \Lambda^2/k^2+i\pi\bigr)}\,.
\end{equation}
We can obtain the scattering amplitude by taking the inverse
Fourier transform and find
\begin{equation}
\label{eq:16}
\psi^{(+)}(\vec{r})=e^{ikx}-2m\lambda\,\psi^{(+)}(\vec{r}=0)
{i\over 4}H_0^{(1)}(kr)\,\,,
\end{equation}
where 
\begin{equation}
{i\over 4}H_0^{(1)}(kr)= \! \int \! {d^2p\over
(2\pi)^2}e^{i\vec{p}\cdot\vec{r}}{1\over
\vec{p}^2-\vec{k}^2+i\epsilon}
\stackrel{r\rightarrow\infty}{\longrightarrow}{1\over 2\sqrt{2\pi
kr}}e^{i(kr+{\pi\over 4})}
\end{equation}
is the two-dimensional Green's function. If we compare
Eq.~(\ref{eq:16}) with the asymptotic form Eq.~(\ref{eq:asy}), we
identify the scattering amplitude as
\begin{equation}
f(\theta)=-{1\over \sqrt{2\pi k}}{1\over {1\over m\lambda}+{1\over
2\pi}\log (\Lambda^2/k^2)+i/2 }\,\,. \label{eq:scat}
\end{equation}
To eliminate the dependence on the cutoff, we note that the 
scattering amplitude has a pole at 
\begin{equation}
E_{\rm bs}=-{\Lambda^2\over 2m} e^{2\pi/\lambda m}\,,
\end{equation}
indicating the presence of a (single) bound state. Using this
binding energy as a parameter, we can rewrite Eq.~(\ref{eq:scat})
as
\begin{equation}
f(\theta)=\sqrt{2\over \pi k}{\pi \over \log ({k^2 \over -2mE_{\rm
bs}})-i\pi} \,,
\end{equation}
which is now expressed only in terms of experimental
quantities. Equivalently, we can characterize the scattering
amplitude in terms of a phase shift 
\begin{equation}
\delta_n(k)=\delta_{n,0}\,\cot^{-1}\bigl({1\over \pi}\bigr)
\log \bigl({k^2\over-2mE_{\rm bs}}\bigr)\,,
\end{equation}
or in terms of the differential
scattering cross section
\begin{equation}
{d\sigma\over d\Omega}={2\over \pi k}{1\over 1+
\pi^{-2}\log^2\bigl({k^2\over -2mE_{\rm bs}}\bigr)}\,.
\end{equation}
The existence of this energy
dependent phase in the
$n=0$ channel or of the fixed energy bound state is clear evidence
of anomalous symmetry breaking. 

The manifestation of anomalous symmetry breaking in quantum
mechanics for the two-dimensional delta function interaction has
been explored previously$^{\ref{bib:del}}$ and was reviewed to
set the context for our primary topic --- the 
$1/r^2$ potential. As we shall see, there is an anomaly in this
case also, and although there are similarities to the
$\delta^2(\vec{r})$ case, there are also important differences, one
of which is the fact that there are experimental consequences.

\subsection{$1/r^2$ Potential}\label{sec:r2}
Consider the potential 
\begin{equation}
V(r)={\lambda\over r^2} 
\end{equation}
in three spatial dimensions. We define the partial wave amplitude
$R_\ell(r)\equiv u_\ell(r)/r$ and find
\begin{equation}\
\label{eq:sch1}
\left[-{d^2\over dr^2}+{\ell(\ell+1)+2m\lambda\over
r^2}-k^2\right]R_\ell(r)=0 \,,
\end{equation}
so that invariance under the scale transformation $r\rightarrow\mu
r,\,k\rightarrow k/\mu$ again holds. It is clear that the solutions 
to Eq.~(\ref{eq:sch1}) can be written in terms of Bessel functions
\begin{equation}
u_\ell(r)\sim \sqrt{kr}J_{\rho+{1\over
2}}(kr)\quad{\rm and}\quad \sqrt{kr}N_{\rho+{1\over 2}}(kr),
\end{equation}
where we have defined 
\begin{equation}
\rho(\rho+1)\equiv \ell(\ell+1)+2m\lambda \,.
\end{equation}
Because $\rho$ is defined via a quadratic equation, it is
apparent that the character of the solutions must change when the
discriminant
\begin{equation}
D_\ell=(\ell+{1\over 2})^2+2m\lambda \label{eq:dis}
\end{equation}
becomes negative. For
repulsive interactions, $\lambda>0$, there is nothing unusual. In
general, the
orders of the Bessel functions become irrational, but the problem
can be solved as usual. Because the
wavefunction must be regular at the origin, we must have
\begin{equation}
u_\ell(r)/r\propto {1\over \sqrt{kr}}J_{\rho+{1\over 2}}(kr)
\stackrel{r\rightarrow\infty}{\longrightarrow}{1\over
kr}\sin(kr-\rho{\pi\over 2}) \,,
\end{equation}
and the scattering phase shift can be read off as
\begin{equation}
\delta_\ell=\left(\ell+{1\over 2}-\sqrt{(\ell+{1\over
2})^2+2m\lambda}\right){\pi\over 2} \,.\label{eq:non}
\end{equation}
Because $\delta_\ell$ is independent of $k$, this result is
consistent with the expected scale invariance.

Now consider the case of an attractive
potential, $\lambda<0.$ As long as the discriminant is positive,
things go through as before. However, once $D_\ell<0$, the
potential has overcome the centrifugal barrier, and the
order $\rho$ of the Bessel function becomes
imaginary.$^{\ref{bib:arg}}$ First consider the case of a
bound state, in which case the boundary condition as
$r\rightarrow\infty$ demands that the solution be constructed in
terms of the Bessel function
$K_{i\Xi_\ell}(\mu r)$, where we have defined $\sqrt{D_\ell}\equiv
i\Xi_\ell$ and the binding energy as
\begin{equation}
E_{\rm bs}\equiv -{\mu^2\over 2m} \,.
\end{equation}
The asymptotic
behavior is then
\begin{equation} u_\ell(r)\propto \sqrt{\mu r}K_{i\Xi_\ell}(\mu
r)\stackrel{r\rightarrow\infty}{\longrightarrow}\sqrt{\pi\over 2}e^{-\mu
r}\, .
\end{equation} 
The allowed values of $\mu$ (and thereby of the binding
energy) are determined by the boundary condition that the
wavefunction vanish at the origin. From the behavior
\begin{equation}
K_{i\Xi_\ell}(\mu
r)\stackrel{r\rightarrow 0}{\longrightarrow}-\sqrt{\pi\over
\Xi_\ell\sinh(\pi\Xi_\ell)} \sin\left [\Xi_\ell\log\left({\mu
r\over 2}\right)-{\rm arg}[\Gamma(1+i\Xi_\ell)]\right ] \,,
\end{equation}
where {\rm arg} indicates the phase of the following complex
number, we see that the wavefunction goes through infinitely
many zeroes as
$r\rightarrow 0$. As a consequence, the spectrum becomes
continuous and unbounded from below, implying that, despite its 
appearance, the Hamiltonian is not self-adjoint. Similarly, the
phase shift of the scattering wavefunction as
$r\rightarrow 0$ is undetermined.

One mathematically attractive
solution to these problems is to define a so-called self-adjoint
extension of the Hamiltonian by specifying a particular
boundary condition at $r=0$.$^{\ref{bib:rjk}}$ For example,
each member of the continuum of self-adjoint extensions of this
Hamiltonian can be characterized by the (energy-independent)
scattering phase shift as
$r\rightarrow 0$.$^{\ref{bib:PP}}$ We briefly discuss
self-adjoint extensions in the applications of the $1/r^2$
potential, but this regularization method does not directly
illustrate anomalous symmetry breaking.

We instead opt for an alternative route and introduce a
short distance cutoff $a$ and
demand that the wavefunction vanish at this
point, $u_\ell(r=a)=0$, and that the physics be independent of the
choice of cutoff parameter.$^{\ref{bib:arg}}$ This
prescription yields the points
\begin{equation}
\mu_n a=2\exp \bigl({{\rm
arg}[\Gamma(1+i\Xi_\ell)]-n\pi\over
\Xi_\ell}\bigr) \,.
\end{equation}
In order that $\mu_n a\rightarrow 0$, we require $\Xi_\ell<<1$ and,
because ${\rm arg}[\Gamma(1+i\Xi_\ell)]=-\gamma\Xi_\ell+{\cal
O}(\Xi_\ell)^2$, the energy becomes 
\begin{equation}
E_{n,\ell}=-{1\over 2m}\left({2e^{-\gamma}\over a}\right)^2 e^{-2\pi n/
\Xi_\ell} \,.\label{eq:exp}
\end{equation}
In order that the ground
state ($n=1,\,\ell=0$) energy remains finite and well-defined as
$a\rightarrow 0$, we require
$\Xi_{\rm gs}=\Xi_{\rm gs}(a)\rightarrow 0^+$, which demands the
scaling behavior \begin{equation} {2\pi\over \Xi_{\rm
gs}(a)}=-2\log\left({\mu a\over 2}\right)-2\gamma \,.
\end{equation} This relation does not predict the value of $\mu$,
but rather defines the scaling of $\Xi_{\rm gs}(a)$ as
$a\rightarrow 0$ in terms of the {\it experimental} value of
$\mu=\sqrt{2mE_{\rm gs}}$. The corresponding ground state
wavefunction is
\begin{equation} \Psi_{\rm gs}(r)={\mu\over \sqrt{2\pi r}}K_0(\mu
r) \,.
\end{equation}
The very existence of a bound state implies the
presence of an energy scale and the breaking of scale
invariance as a result of quantization, just as in the
case of the two-dimensional delta function --- a quantum mechanical
example of an anomaly!

Another similarity with the two-dimensional delta function potential
is that there is but a single bound state. This similarity can be
seen from the fact that because $\Xi_{\rm gs}\rightarrow 0^+$ and
$D_{\ell}>D_0$, the discriminant in any but the $s$-wave channel
must be positive so that no anomaly (and no bound state) can occur.
Similarly in the
$\ell=0$ case but with $n>1$, we see
from Eq.~(\ref{eq:exp}) that such states cannot have nonzero
binding energies, because
\begin{equation}
{E_{n,0}\over E_{1,0}}=e^{-2\pi(n-1)/\Xi_{\rm gs}}
\stackrel{\Xi_{\rm gs}\rightarrow 0^+}{\longrightarrow}0
\,.
\end{equation}
We summarize this discussion with the observation that from
Eq.~(\ref{eq:dis}) and the following, there exists a critical value
of the coupling constant, $2m\lambda=-{1\over 4}$, below which
there exists a single bound state and above which there are no
bound states.

We can also consider scattering in the presence of
an anomaly.
In this case the solutions of the partial wave equation
must be linear combinations of $H_\rho^{(1)}(kr)$ and
$H_\rho^{(2)}(kr)$:
\begin{equation}
u_\ell(r)/\sqrt{r}=A_1H_\rho^{(1)}(kr)+A_2H_\rho^{(2)}(kr) .
\end{equation}
{}From the asymptotic dependence
\begin{equation}
u_\ell(r)/\sqrt{r}\stackrel{r\rightarrow\infty}{\longrightarrow}
\sqrt{2\over \pi kr}\left(A_1e^{i(kr-{1\over 2}\rho\pi-{\pi\over 4})}
+A_2e^{-i(kr-{1\over 2}\rho\pi-{\pi\over 4})}\right)\, ,
\end{equation}
then for the $s$-wave channel in which $\Xi_{0}\rightarrow 0^+$, we
identify the scattering phase shift via
\begin{equation}
{A_1\over A_2}= e^{i(2\delta_0-{\pi\over 2})} \,.\label{eq:fund}
\end{equation}
On the other hand, for small values of $r$, we find
\begin{eqnarray}
u_0(r)/\sqrt{r}&=&A_1(J_{i\Xi_0}(kr)+iN_{i\Xi_0}(kr))+
A_2(J_{i\Xi_0}(kr)-iN_{i\Xi_0}(kr))\nonumber\\
&&\stackrel{r\rightarrow 0}{\longrightarrow}A_1\left({({1\over
2}kr)^{i\Xi_0}\over \Gamma(1+i\Xi_0)}(1+{\cosh\pi\Xi_0\over
\sinh\pi\Xi_0})-{1\over \sinh\pi\Xi_0}{({1\over
2}kr)^{-i\Xi_0}\over \Gamma(1-i\Xi_0)}\right)\nonumber\\
&+&A_2\left({({1\over
2}kr)^{i\Xi_0}\over \Gamma(1+i\Xi_0)}(1-{\cosh\pi\Xi_0\over
\sinh\pi\Xi_0})+{1\over \sinh\pi\Xi_0}{({1\over
2}kr)^{-i\Xi_0}\over \Gamma(1-i\Xi_0)}\right) \,.
\end{eqnarray}
The requirement that $u_0(r=a)=0$
yields
\begin{eqnarray}
{A_1\over A_2}&=&{e^{i\sigma}(1+i\xi)e^{-\pi\Xi_0}-
e^{-i\sigma}(1-i\xi)\over e^{i\sigma}(1+i\xi)e^{\pi\Xi_0}-
e^{-i\sigma}(1-i\xi)}\nonumber\\
&=&{\xi+\tan\sigma+i\tanh{1\over 2}\pi\Xi_0(1-\xi\tan\sigma)\over 
\xi+\tan\sigma-i\tanh{1\over 2}\pi\Xi_0(1-\xi\tan\sigma)} \,,
\end{eqnarray}
where we have defined
\begin{equation}
\label{eq:comp}
i\xi={\Gamma(1-i\Xi_0)-\Gamma(1+i\Xi_0)\over
\Gamma(1-i\Xi_0)+\Gamma(1+i\Xi_0)}\quad {\rm and} \quad e^{i\sigma}=
({1\over 2}ka)^{i\Xi_0}=\exp(i\Xi_0\log{1\over 2}ka)\, .
\end{equation}
If we compare Eq.~(\ref{eq:comp}) with Eq.~(\ref{eq:fund}), we
can identify
\begin{equation}
\delta_0(k)-{\pi\over 4}=\tan^{-1}{\tanh{1\over
2}\pi\Xi_0(1-\xi\tan\sigma)\over \xi+\tan\sigma} \,.
\end{equation}
In the limit that $\Xi_0\rightarrow 0^+$, we have then
\begin{equation}
\tan(\delta_0(k)-{\pi\over 4})\approx {1\over
2}\pi\Xi_0{1-\gamma\Xi_0\log{1\over 2}ka\over
\gamma\Xi_0+\tan\Xi_0\log{1\over 2}ka} \,.
\end{equation}
Using the scaling behavior
\begin{equation}
\Xi_0(a)={-\pi\over \log{1\over 2}\mu
a}+\gamma \,,
\end{equation}
we find that
\begin{equation}
\tan(\delta_0(k)-{\pi\over 4})={1-\cot\delta_0(k)\over
1+\cot\delta_0(k)}
\approx{\pi\over \log{k^2\over \mu^2}} \,,
\end{equation}
which yields the partial wave amplitude
\begin{equation}
ka_0(k)={1\over \cot\delta_0(k)-i}={1\over \left({\log{k^2\over
\mu^2}-\pi
\over \log{k^2\over \mu^2}+\pi}\right)-i} \,.\label{eq:pol}
\end{equation}
As a check we verify that that Eq.~(\ref{eq:pol}) has a pole at the
bound state energy $k^2=2mE_{\rm bs}=-2m\mu^2$.

Because
\begin{equation}
\Xi_\ell=\sqrt{(\ell+{1\over
2})^2-2m|\lambda|}>\Xi_0=\sqrt{{1\over 4}-2m|\lambda|}\rightarrow
0 \,,
\end{equation}
we see that there is no anomaly in the non-$s$ wave
states so that the scattering phase shift is given by its
non-anomalous value given in Eq.~(\ref{eq:non}). The form of the
scattering amplitude is
\begin{eqnarray}
f(\theta)&=&{1\over k}\sum_{\ell=1}^\infty(2\ell+1)i^\ell
\exp(i\delta_\ell\sin\delta_\ell) P_\ell(\cos\theta)\nonumber\\
{}&&+{1\over k}{1\over \left({\log{k^2\over
\mu^2}-\pi
\over \log{k^2\over \mu^2}+\pi}\right)-i} \,,
\end{eqnarray}
where $\delta_\ell$ is given by Eq.~(\ref{eq:non}).

\subsection{An Aside}
Before proceeding to applications of this formalism, it is
interesting (but unrelated to considerations of the anomaly) that
there is another curious feature of the $1/r^2$ interaction, as
previously pointed out by Kayser.$^{\ref{bib:kay}}$ If one
solves for the classical scattering angle for motion in the
presence of such a potential, the result is
\begin{equation}
\theta_{\rm cl}=\pi\left(1-{L\over
\sqrt{L^2+2m\lambda}}\right) \,,\label{eq:cl}
\end{equation}
where $L$ is the angular momentum. Taking the interaction to be repulsive 
($\lambda>0$) and solving for the cross section, we find
\begin{eqnarray}
{d\sigma_{\rm cl}\over d\Omega}&=&{1\over p^2}{L\over
\sin\theta}\left|{dL\over d\theta}\right|\nonumber\\
&=&{\lambda\over 2\pi E}{1-x\over x^2(2-x)^2\sin\pi x} \,,
\end{eqnarray}
where $p=\sqrt{2mE}$ is the incoming momentum, and we have defined
$x=\theta/\pi$. Thus the cross section is {\it linear} in the
coupling constant, in apparent violation of the simple Born
approximation result when the potential becomes weak. In
Ref.\cite{kay} it is shown that the same cross section results from
a quantum mechanical evaluation. The resolution of the apparent
paradox lies in the fact that the classical scattering condition
in Eq.~(\ref{eq:cl}) can be satisfied only when $2m\lambda>>1$.
This condition can be seen from the result
$L\theta>>1$ which implies that 
\begin{equation}
2m\lambda>>{1\over \pi^2}{2-x\over x(1-x)^2} \,.
\end{equation}
Thus the classical result corresponds to the strong coupling regime,
where the Born approximation is not appropriate.

\section{Applications}\label{sec:apply}
A remarkable feature of the above analysis of the $1/r^2$ potential
is that this interaction is realized in nature. As has
recently been emphasized, one such application is to that of a 
charge interacting with a point dipole.$^{\ref{bib:fan}}$ 
Because the potential outside an
electric dipole $\vec{p}$ is given by
\begin{equation}
\phi(\vec{r})={\vec{p}\cdot\vec{r}\over 4\pi r^3} \,,
\end{equation}
the potential energy for a charge $e$ in the vicinity is given by
\begin{equation}
V(\vec{r})={e\vec{p}\cdot\vec{r}\over 4\pi r^3} =
{\sigma\cos\theta\over r^2} \,,
\end{equation}
which has the desired $1/r^2$-dependence. Point dipoles are not
physical, but a good approximation is provided by a polar molecule
such as water and the point charge can be taken to be a nearby
electron. To analyze this system by means of the formalism
developed in Sec.~2, we write the bound state solution as in
Ref.\cite{fan}:
\begin{equation}
\psi(\vec{r})={1\over r}u(r)\Theta(\theta) \,.
\end{equation}
Then the equations determining the radial and angular dependence are
given by
\begin{eqnarray}
\left(-{1\over 2m}{d^2\over dr^2}+{\gamma\over 2m
r^2}\right)u(r)&=&Eu(r)\nonumber\\
(\hat{L}^2+2m\sigma\cos\theta)\Theta(\theta)&=&
\gamma\Theta(\theta) \,,\label{eqn:eig}
\end{eqnarray}
and the separation constant $\gamma$ is related to the actual
coupling
$\sigma$ of the point-dipole potential by Eq.~(\ref{eqn:eig}).
If we
use the normalized Legendre polynomials 
\begin{equation}
\sqrt{(2\ell+1)\over 2}P_\ell(\cos\theta) \,,
\end{equation}
as a basis, Eq.~(\ref{eqn:eig}b) can be written as a matrix
equation
\begin{equation}
M_{\ell\ell'}\Theta_{\ell'}=0 \,,
\end{equation}
with
\begin{eqnarray}
\label{eq:matrix}
M_{\ell\ell'}&=&\delta_{\ell\ell'}(\ell(\ell+1)-\gamma) \nonumber
\\ {}&& +\, 2m\sigma\left[{\ell\over
\sqrt{(2\ell-1)(2\ell+1)}}\delta_{\ell-1,\ell'}+{\ell+1\over 
\sqrt{(2\ell+1)(2\ell+3)}}\delta_{\ell+1,\ell'}\right] \,.
\end{eqnarray}
Equation~(\ref{eq:matrix}) is an eigenvalue equation, for which the
existence of a solution requires that
\begin{equation}
\label{eq:sol}
{\rm det}M_{\ell\ell'}={\rm det}\left|\begin{array}{cccc}
-\gamma&{2m\sigma\over \sqrt{3}}&0&\ldots\\
{2m\sigma\over \sqrt{3}}&2-\gamma&{4m\sigma\over \sqrt{15}}&\ldots\\
0&{4m\sigma\over \sqrt{15}}&\ldots\\
\ldots&\ldots&\ldots&\ldots\end{array}\right|=0 \,.\label{eq:dip}
\end{equation}
Equation~(\ref{eq:sol}) can be easily solved numerically by
successively evaluating the
$n\times n$ determinant, which converges extremely rapidly:
\begin{eqnarray} n&=&2:\quad (2m\sigma)^2=-3\gamma(2-\gamma)\nonumber\\
n&=&3:\quad (2m\sigma)^2=-3\gamma(2-\gamma)\left({1\over 1-{4\gamma\over
5(6-\gamma)}}\right) . \label{eq:ite}
\end{eqnarray}
{}From our study of the
$1/r^2$ potential in Sec.~\ref{sec:r2}, we know that there
exists a critical value of the coupling constant,
$2m\lambda=-{1\over 4}$, below which there exists a bound state
and above which there does not. Hence from Eq.~(\ref{eq:dip}), we
observe that there must exist a critical value of the dipole moment
$\sigma^*$, defined via
\begin{equation}
{\rm det}\,M_{\ell\ell'}(\gamma=-{1\over 4},2m\sigma^*)=0 \,,
\end{equation}
such that a bound electron-polar molecule state, an anion, can
(cannot) exist for values of the dipole moment larger (smaller) 
than this critical value. From Eq.~(\ref{eq:ite}), we find
$2m\sigma^*=1.279\ldots$, that is, 
\begin{equation}
p^*={4\pi\over e}\sigma^*=0.640ea_0 \,,
\end{equation}
where $a_0=1/m\alpha$ is the Bohr radius.

This prediction can be checked experimentally, and studies
have been conducted attempting to measure the binding energies of
anion systems.$^{\ref{bib:stu}}$ It has indeed been found
that a minimum dipole moment exists, as shown in Fig.~1;
there does seem to exist a minimum dipole moment $p^*_{\rm
exp}\simeq 0.86ea_0$, as would be expected from the above
considerations. Note that there is a bit of a subtlety here in
that a real molecule is {\it not} a point dipole. Rather the
potential for a finite dipole can be written as
\begin{equation}
V(\vec{r})={qe\over 4\pi}\left({1\over R_+}-{1\over
R_-}\right)={ep\cos\theta\over 4\pi r^2}+U_{\rm br}(\vec{r}) \,, 
\end{equation}
which is the superposition of a pure point dipole interaction with
a scale symmetry-breaking interaction $U_{\rm br}(\vec{r})$.
Because experimentally such an explicitly symmetry breaking term
cannot be eliminated, the observed critical moment is due to a
combination of the anomalous and explicit symmetry breaking
effects. It is clear that the presence of explicit breaking does
not affect the existence of a critical moment, as expected from
the quantum mechanical anomaly. However, the size of $p^*$ {\it
is} affected, as can be seen from the comparison of experimental
and theoretical sizes of the critical moment. This difference is
perhaps not surprising, because in QCD, the combination of
explicit and anomalous symmetry breaking results in a value for the
$\pi^0\rightarrow\gamma\gamma$ decay rate very near that
predicted by the anomaly.$^{\ref{bib:pi}}$

It is interesting that a second application of the $1/r^2$ potential has
recently been discussed. In this case it involves the interpretation of
an experiment involving the interaction of a neutral but polarizable atom
with a charged wire.$^{\ref{bib:sch}}$ Because the electric
field generated by the wire falls off linearly with distance, the
induced electric dipole moment has the form \begin{equation}
\vec{p}=4\pi\alpha_E\vec{E}\propto {\alpha_E\over r} \,,
\end{equation} where $\alpha_E$ is the electric polarizability. 
The corresponding interaction potential is
\begin{equation} U(r)=-{1\over 2}\vec{p}\cdot\vec{E}\propto
-{\alpha_E\over r^2} \,, \end{equation} and falls off as $1/r^2$.
The issue here is not the experimental appearance of a critical
value of the coupling constant because the attractive $1/r^2$
potential is always critical in this cylindrical {\em
two-dimensional} problem,$^{\ref{bib:arg}}$ no matter what the
strength of the electric field produced by the charged wire. The
intriguing aspect of this experiment is that the atoms are
observed to disappear from the system with an absorption cross
section which can be fitted by a classical
argument.$^{\ref{bib:sch}}$ An analogous treatment of the
two-dimensional $1/r^2$ potential, regularized and renormalized as
done here, would have only bound state and elastic scattering
solutions, and no absorption. Bawin and Coon utilized a method
suggested by Radin$^{\ref{bib:sad}}$ to sum over the infinite number
of elastic scattering solutions of the self-adjoint extensions of
the
$1/r^2$ interaction to obtain a quantum mechanical expression that
displayed absorption.$^{\ref{bib:baw}}$ As the parameters of
this initial experiment corresponded to the classical limit
established by Kayser,$^{\ref{bib:kay}}$ the classical limit
of their quantum mechanical treatment recovered the experimental
situation. However, further discussion is beyond the scope of this
paper.

\section{Conclusions}\label{sec:concl}
The phenomenon of symmetry breaking is universal in physics, but
although explicit and spontaneous violations are well-known and are
generally treated in introductory courses, the same cannot be said
of quantum mechanical or anomalous symmetry breaking, where a
symmetry is valid at the classical level but is violated due to
quantization of the theory. This omission is presumably due to the
fact that 
most familiar manifestations of the anomaly, for example, the
two-photon decay of the neutral pion, occur in the context of
quantum field theory and belong in the realm of elementary particle
physics. However, this need not be the case. We
have argued here that one can present this subject within the
realm of ordinary quantum mechanics by studying the violation of
scale symmetry, $r\rightarrow\mu r$, $k\rightarrow k/\mu$. In
earlier papers in this journal, the case of a two-dimensional delta
function was examined.$^{\ref{bib:del}}$ However, this case
is purely an intellectual exercise and has no experimental
consequences. In this paper, we have examined the $1/r^2$
potential, which is anomalous and which {\it does} have
experimental ramifications in the existence of a critical dipole
moment allowing the binding of anions in atomic
physics$^{\ref{bib:stu}}$ and in the recent study of the
interaction of a polarizable atom with a charged
wire.$^{\ref{bib:sch}}$

Finally, it has long been known
that the main features of the quantum mechanical three-body bound
state and some striking aspects of nuclear three-body scattering
are due to an effective $1/r^2$ potential, built from the relative
distances between the particles, including mass factors where
appropriate.$^{\ref{bib:efimov}}$ This effective potential is
displayed most clearly in recent effective field theory
discussions of three-body systems, where it also exhibits scale
invariance violation,$^{\ref{bib:van}}$ but this discussion
is clearly beyond the level of our paper. In any case we believe
that the above discussion is at a level that is appropriate in an
advanced quantum mechanics course, and it is hoped that paper note
will encourage the introduction of this fascinating topic into
such a venue.

\section*{Acknowledgement}
It is a pleasure to acknowledge very interesting discussions with
Boris Kayser and with Michel Bawin about this material. This work
was supported in part by the National Science Foundation
under award NSF-PHY-98-01875 and NSF-PHY-00-70938.

\newpage

\begin{figure}
\begin{center}
\includegraphics*[width=7.5cm]{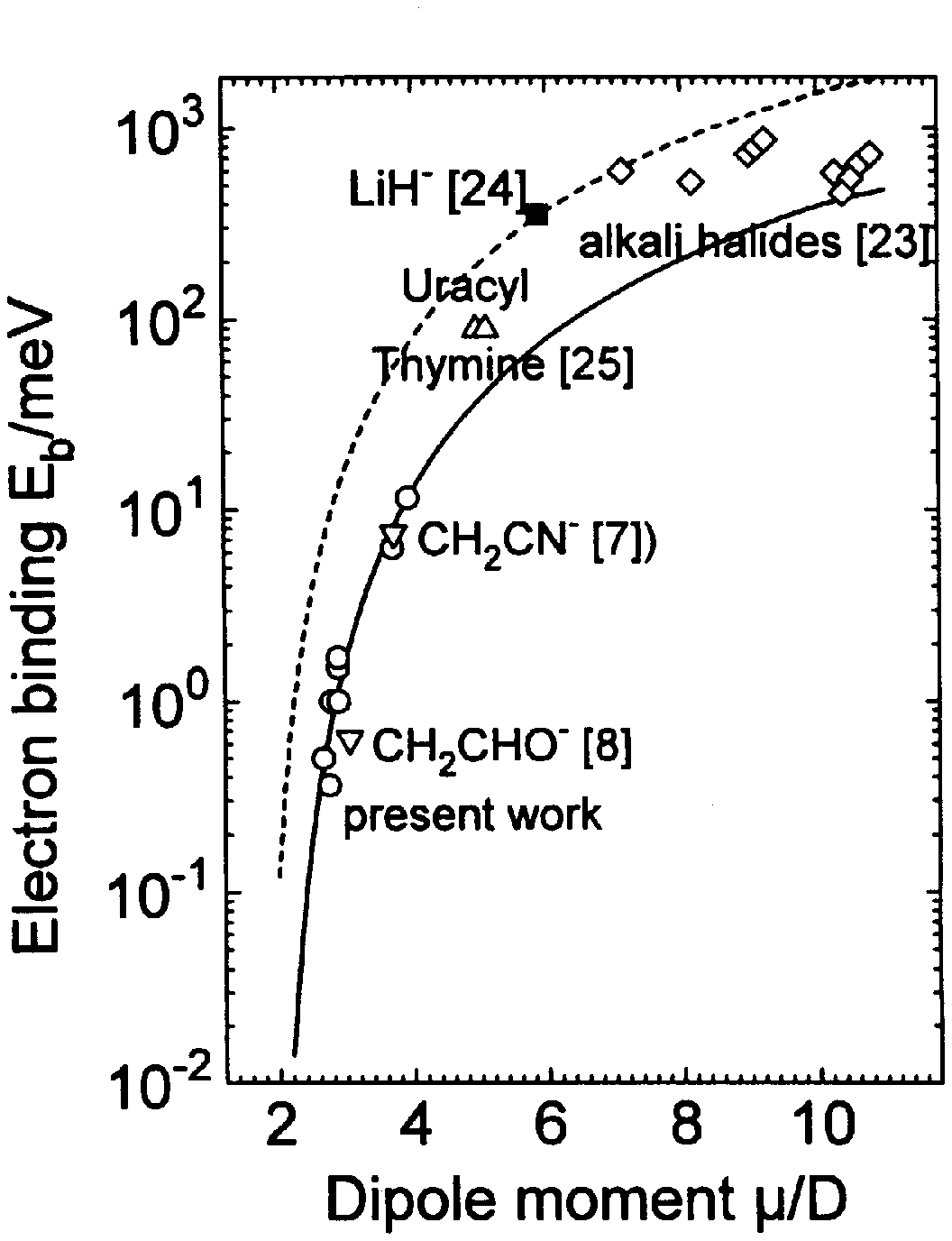}
\caption{Shown is data reported in Ref.\cite{stu} on experimental
anion binding energies. Note that there seems to exist a limiting
value 
$p_{\rm crit}\simeq 2D$, where $D=1$ Debye $\equiv 1\times
10^{-{18}}$\,esu\,cm. Because $ea_0\simeq 2.4$ D, we have $p_{\rm
crit}\simeq 0.86 ea_0$.}
\end{center}
\end{figure}

\end{document}